\documentclass[10pt,reqno,oneside]{amsart}

\usepackage[normalem]{ulem}
\usepackage{amssymb}
\usepackage[pdftex]{color}
\usepackage{hyperref}
\hypersetup{
	colorlinks   = true, 
	urlcolor     = brown, 
	linkcolor    = black, 
	citecolor   = black 
}
\usepackage{array}
\usepackage{mathtools}
\usepackage{multicol}
\usepackage[nameinlink, capitalise, noabbrev]{cleveref}
\usepackage{url}
\usepackage{amsmath}
\usepackage{graphicx,indentfirst,amsmath,amsfonts,amssymb,amsthm,newlfont}
\usepackage{epsfig}
\usepackage{enumitem}
\usepackage{tikzsymbols}
\usepackage{float}
\usepackage[font=footnotesize]{caption} 
\usepackage{subcaption}
\usepackage[a4paper, total={5.6in, 8.5in}]{geometry}

\usepackage[T1]{fontenc}

\newtheorem{theorem}{Theorem}[section]
\newtheorem{corollary}{Corollary}[section]
\newtheorem{proposition}{Proposition}[section]
\newtheorem{definition}{Definition}[section]
\newtheorem{lemma}[theorem]{Lemma}
\newtheorem{remark}{Remark}[section]
\newtheorem{example}{Example}[section]

\newcommand{\B}{\mathcal B}
\newcommand{\M}{\mathcal M}
\newcommand{\R}{\mathbb R}

\newcommand{\Ro}{\mathcal{R}_0}
\newcommand{\F}{\mathcal F}
\newcommand{\E}{\mathcal E}
\newcommand{\K}{\textbf{K}}
\newcommand{\Nt}{\widetilde{N}}
\newcommand{\diag}{\mathrm{diag}}

\begin{document}
	
	\title{Final size of a structured SIRD Model with active-population force of infection}

\author{Alison M.V.D.L. Melo}
\address{A.M.V.D.L.M.: Universidade Federal do Vale do São Francisco - UNIVASF, 56304-917, Petrolina, Brazil}
\email{alison.melo@univasf.edu.br}

\author{Matheus C. Santos}
\address{M.C.S.: Departamento de Matem\'atica Pura e Aplicada-- IME, Universidade Federal do Rio Grande do Sul - UFRGS, 91509-900, Porto Alegre, Brazil}
\email{matheus.santos@ufrgs.br}

	\begin{abstract}
		We consider a SIRD epidemic model for a population composed of two groups of individuals with asymmetric interaction where the force of infection depends on the active (alive) population in each group, rather than on the total population as in the classical formulation. We prove that the final state for susceptible individuals is always positive and characterize it as the unique fixed point of a map. We also relate the final size to the basic reproduction number and show that the final number of susceptibles decreases when transmission rates increase. Numerical simulations compare the active-population and classical two-group SIRD models, showing differences in final size and the occurrence of multiple epidemic waves. The convergence of the fixed point approach is also illustrated.\end{abstract}

	{\thispagestyle{empty}} 

	\maketitle

	\section{Introduction}

	\medskip
	
	\everymath{\displaystyle}

In this paper, we consider a two-group SIRD model to analyze a population divided into two interacting groups, where infected individuals from one group can transmit the disease to susceptibles in either group, but with potentially different transmission coefficients. The model also includes group-specific recovery and mortality rates and infected individuals may recover or die due to the disease. The main difference from standard models is that deceased individuals are removed from the pool of active individuals. 

We will denote by $S_i(t)$, $I_i(t)$, $R_i(t)$, and $D_i(t)$ the numbers of susceptible, infectious, recovered, and deceased individuals in group $i$, respectively.

The system of equations for the two-group model studied in this work is the following SIRD model with active population in the force of infection
	\begin{equation}\label{Eq.SIRDm}
		\left\{ \begin{array}{lll}
			S_1'  = -\left(\beta_{11}\frac{I_1}{N_1} +\beta_{12}\frac{I_2}{N_2} \right) S_1    &,& S_2'  = -\left(\beta_{21}\frac{I_1}{N_1} +\beta_{22}\frac{I_2}{N_2} \right) S_2 \\
			I_1'  =\left(\beta_{11}\frac{I_1}{N_1} +\beta_{12}\frac{I_2}{N_2} \right) S_1 - (\gamma_1+\mu_1)I_1  &,& I_2'  =\left(\beta_{21}\frac{I_1}{N_1} +\beta_{22}\frac{I_2}{N_2} \right) S_2  - (\gamma_2+\mu_2)I_2 \\
			R_1'  = \gamma_1 I_1 &,&R_2'  = \gamma_2 I_2 \\
			D_1'  = \mu_1 I_1 &,&D_2'  = \mu_2 I_2 
		\end{array}\right.
	\end{equation}
	with initial conditions
	$S_i(0)=S_{i0} ,\; I_i(0)=I_{i0},\; R_i(0)=R_{i0},\;D_i(0)=D_{i0}\in \R_+$ and where $N_i(t)=S_i(t)+I_i(t)+R_i(t)$ are the active (alive) individuals in group $i$ at time $t$, for $i=1,2$. The parameters $\beta_{i,j},\alpha_i,\gamma_i,\mu_i$ are positive. We will refer to this model simply as \textit{active population SIRD} model. Let us write $\Nt_i = N_i(t)+D_i(t)$ the constant value of the total number of individuals (alive and deceased) in each group. 

Models with structured populations, such as \eqref{Eq.SIRDm}, partition individuals into groups that may differ in susceptibility, infectiousness, social behavior, as well as in recovery and mortality rates. Such multigroup models are particularly relevant when distinct demographic or behavioral subpopulations are present. For example, groups with unequal access to health care or age-structured groups that respond differently to a given disease. Moreover, these models can be adapted to incorporate mobility phenomena, allowing individuals to move temporarily or permanently between groups over the course of the epidemic.

Classical compartmental models assume that the force of infection in a given group $i$ is proportional to $I_i/\Nt_i$, where $I_i$ denotes the number of infectious individuals and $\Nt_i$ the total population size, including both active and removed individuals, such as deceased ones. However, this formulation implicitly accounts for individuals who are no longer susceptible or even alive, which can underestimate transmission dynamics in scenarios with significant mortality or long-term removal. A more realistic approach considers the force of infection proportional to $I_i/N_i$, where $N_i$ represents only the active (alive) population in group $i$. This feature alters the force of infection in a nontrivial way, as the effective contact rate is modulated by the dynamic size of the active population, introducing an additional nonlinearity into the system that alters both transient dynamics and equilibrium behavior, especially in populations with high-mortality rates.

These multigroup models provide a framework for quantifying the spread of infectious agents through the population, the effectiveness of interventions, and the ultimate reach of an epidemic, often called its final size. This quantity represents the number (or fraction) of individuals who escape infection after the epidemic concludes and is defined for each group $i$ as $S_i^\infty = \lim_{t \to \infty}S_i(t)$. The study of final epidemic size is already known for many epidemic models, especially in the absence of disease-induced mortality (\cite{Andreasen, Magal1, Magal2, MKS, Arino, MaEarn, Miller}). These works typically rely on the invariance of certain conserved quantities, system of integro-differential equations, the use of the next generation matrix and even the underlying stochastic process. However, models with mortality and time-varying population size remain less explored in this context. We adapt some of these techniques to present a first integral to the system \eqref{Eq.SIRDm} and to show that the final sizes of both groups are strictly positive and satisfy a system of nonlinear equations derived from integrating the governing ODEs.

The main result of our work is the characterization of $S^\infty := (S_1^\infty,S_2^\infty)$ as a fixed point of a nonlinear map $T:\mathbb{R}_+^2 \to \mathbb{R}_+^2$, leading to a closed-form expression linking the final size to the initial conditions and model parameters. We analyze this map and prove the existence and uniqueness of the fixed point under suitable assumptions on the model parameters. In particular, we show that under hypotheses ensuring subhomogeneity or spectral conditions on the Jacobian of $T$, the fixed point iteration converges monotonically to the unique solution representing the final susceptible populations. Thus, this formulation provides not only an implicit description of the final epidemic size but also a practical iterative method for its computation. 

Furthermore, we connect our findings to the basic reproduction number $\Ro$ by constructing the next generation matrix. We derive a relation between the final susceptible states and $\Ro$ through a logarithmic identity involving the next generation matrix and its principal eigenvector.

To illustrate our theoretical results, we present numerical simulations that compare our model with the classical two-group SIRD framework. These experiments reveal qualitative differences in the epidemic dynamics and show how the mortality rate can significantly affect the final size. Overall, the simulations highlight substantial discrepancies in both the transient behavior and the final size predictions across the two models, and we also numerically illustrate the convergence of the final-size formulation given by the associated fixed-point problem.

\section{Hypotheses and basic properties of the model}
In this section we present the main hypotheses of the paper, fix some notation and collect some elementary qualitative properties of system \eqref{Eq.SIRDm} that will be used later.

We will assumne the following hypothesis on the infection and mortality rates of the system:
	
	\begin{enumerate}[label=\textbf{H\arabic*)}]
		\item\label{H1} The parameter matrices $\B$ and $\M$ are positive, where
		\[\B = \begin{pmatrix}
			\beta_{11}& \beta_{12} \\
			\beta_{21}& \beta_{22}
		\end{pmatrix} \quad \mbox{and} \quad\M = \begin{pmatrix}
			\mu_1& 0 \\
			0& \mu_2
		\end{pmatrix}\].
        \end{enumerate}

        Additionally we will assume either one of the following hypotheses:
        
        \begin{enumerate}[label=\textbf{H2\alph*)}]
		\item\label{H2a} $\frac{\beta_{ij}}{\mu_j}\geq1$, for all $i,j\in\{1,2\}$.
		
		\item\label{H2b} $\frac{\beta_{i1}}{\mu_1}+\frac{\beta_{i2}}{\mu_2}<1$, for $i=1,2$.
		
		\item\label{H2c} $\frac{\beta_{i1}}{\gamma_1+\mu_1}\frac{1}{Z_1}+\frac{\beta_{i2}}{\gamma_2+\mu_2}\frac{1}{Z_2}< \frac{1}{S_{i0}}$, for $i=1,2$, where $Z_j=\frac{\gamma_j}{\gamma_j+\mu_j}N_j(0)+\frac{\mu_j}{\gamma_j+\mu_j}R_{j0}$

    	\item\label{H2d} $\frac{\beta_{i1}}{\gamma_1+\mu_1}\frac{S_{10}}{Z_1}+\frac{\beta_{i2}}{\gamma_2+\mu_2}\frac{S_{20}}{Z_2}< 1$, for $i=1,2$.
	\end{enumerate}

From a biological perspective, the hypotheses \ref{H2a} and \ref{H2b} describe different regimes for how strong transmission is compared to recovery and disease-induced death. Condition \ref{H2a} says that, for each source group $j$, an infectious individual typically generates at least one effective infectious contact before dying from the disease, corresponding to a rather aggressive transmission scenario. In contrast, \ref{H2b} expresses that, for each target group $i$, the total infection pressure coming from both groups, when viewed over the disease-mortality time scale in each source group, is relatively modest; in other words, even if infection can spread between groups, the average infectious output per infectious individual remains limited.

From a numerical standpoint, hypothesis \ref{H2a} is satisfied for Ebola virus disease during the 2014 West Africa outbreak. The WHO Ebola Response Team reported case fatality ratios $f$ in Guinea in the interval $66.7-74.3\%$ and basic reproduction numbers $\mathcal{R}_0\in[1.44,2.01]$, which yields $\beta/\mu = \mathcal{R}_0/f\in[1.93,3.01]$ (\cite{WHO2014_Ebola}). Also, for COVID-19 in Wuhan, early estimates indicated $\mathcal{R}_0\in[1.4,3.9]$ (\cite{Li2020_COVID}), while severity analyses placed the infection fatality ratio in $0.39-1.33\%$ (\cite{Verity2020_IFR}). Combining these ranges, we obtain $\beta/\mu\in[3.68,3.93]$. Hypothesis H2b requires that a typical infectious individual, before dying from the disease, generates on average fewer than one effective infection. A plausible example is person-to-person transmission of Nipah virus in Bangladesh: with a case fatality ratio between 40–75\%, an average number of secondary cases $\mathcal{R}_0=0.33$ (\cite{WHO_Nipah_2023,Nikolay}), we can estimate $\beta/\mu\in[0.44,0.83]$. Similarly, for human infection with avian influenza A(H5N1), global reviews report a case fatality risk around 50–60\%(\cite{Lai2016_H5N1}). Household transmission studies in Indonesia estimate $\mathcal{R}_0\in[0.1,0.25]$ (\cite{Aditama2012_H5N1,Bui, Park2018_Transmissibility}). These values imply the ratios $\beta/\mu$ approximately in $[0.17,0.5]$. Although most Nipah virus and A(H5N1) infections arise from animal-to-human spillover, limited human-to-human transmission can occur. Here we focus only on the latter, so $R_0$ is interpreted under human-to-human transmission.

\begin{definition}
We will denote the vectors of inicial conditions of the system \eqref{Eq.SIRDm} by
\[S_0=\begin{pmatrix}
    S_{10}\\ S_{20}
\end{pmatrix}\,,\; 
I_0=\begin{pmatrix}
    I_{10}\\ I_{20}
\end{pmatrix}\,,\; 
R_0=\begin{pmatrix}
    R_{10}\\R_{20}
\end{pmatrix}\,,\; 
D_0=\begin{pmatrix}
    D_{10}\\D_{20}
\end{pmatrix}\,,\; 
\]
and the final state vectors by
\[S^\infty = \begin{pmatrix}
    S^\infty_{1}\\S^\infty_{2}
\end{pmatrix} = \lim_{t\to\infty} \begin{pmatrix}
    S_{1}(t)\\S_{2}(t)
\end{pmatrix},;\qquad N^\infty = \begin{pmatrix}
    N^\infty_{1}\\N^\infty_{2}
\end{pmatrix} = \lim_{t\to\infty} \begin{pmatrix}
    N_{1}(t)\\N_{2}(t)
\end{pmatrix}. \]
\end{definition}

\begin{definition}
 Let us define the following notation for partial ordering of vectors in $\R^2$: given $X=(x_1,x_2),Y=(y_1,y_2)\in \R^2$, we say that
	\[\begin{array}{lll}
		X\leq Y \Longleftrightarrow x_i \leq y_i &\mbox{ for all }&i=1,2\,,\\
		X< Y \Longleftrightarrow  X\leq Y \mbox{ and } x_i < y_i &\mbox{ for some }&i=1,2\,,\\ 
		X\ll Y\Longleftrightarrow x_i < y_i &\mbox{ for all } &i=1,2\,.
	\end{array}
	\]
\end{definition}

The following two results are classical ones for epidemiological models without vital dynamics: forward invariance of the biologically relevant region, global existence of solutions, and extinction of the infectious classes.

\begin{theorem}\label{thm:wellposed}
Assume \ref{H1} and $S_{0},I_{0},R_{0},D_{0}\ge0$, with $S_0+I_0>0$. Then:
\begin{enumerate}[label=\roman*)]
\item System \eqref{Eq.SIRDm} has a unique solution defined for all $t\ge0$.
\item For each fixed pair $\Nt_1,\Nt_2>0 $, the set
\[
\mathcal{D} := \Bigl\{(S,I,R,D)\in\mathbb{R}_+^8:\;
S_i+I_i+R_i+D_i = \Nt_i,\ i=1,2\Bigr\}
\]
is positively invariant. It follows that along any solution with initial data in $\mathcal D$ we have
$0\le S_i(t),I_i(t),R_i(t),D_i(t)\le \Nt_i$ for all $t\ge0$.
\end{enumerate}
\end{theorem}

\begin{theorem}\label{thm:extinction}
Under \ref{H1} and $S_0\gg0$, $I_0\ge0$, every solution of \eqref{Eq.SIRDm} satisfies:

\begin{enumerate}[label=\roman*)]
\item the limit values of all compartments exist, i.e., for $i=1,2$ there exist
\[S_i^\infty = \lim_{t\to\infty}S_i(t),\;\quad I_i^\infty = \lim_{t\to\infty}I_i(t),\;\quad R_i^\infty = \lim_{t\to\infty}R_i(t),\;\quad D_i^\infty = \lim_{t\to\infty}D_i(t)\]
\item $I_1^\infty=0$ and $I_2^\infty=0$. In particular, the only equilibria of \eqref{Eq.SIRDm} is disease-free.
\end{enumerate}
\end{theorem}

\begin{proof}
For the first item, the system \eqref{Eq.SIRDm} implies that the functions $S_i, R_i$ and $D_i$ are monotone. Since Theorem \ref{thm:wellposed} guarantees that they are bounded by 0 and $\Nt_i$, it follows that they converge, as $t\to \infty$. Thus, by the definition of $\Nt_i$, the function $I_i$ also converges when $t\to\infty$ to a value $I^\infty_i$. 

For the second item, we obtain from the equations \eqref{Eq.SIRDm} for $R_i$ and the existence of $R_i^\infty$ that $I_i(t)$ is integrable in $[0,\infty)$. This fact together with the existence of $I_i^\infty$ implies the desired result.
\end{proof}

In the next section, we will show that the values $S^\infty_i$ are always positive and show a way of obtaining it as a fixed point iteration in $\R^2$.

	\section{Final state}

	Let us begin by presenting a relation between the epidemic limit values.
	
	\begin{lemma}\label{Lemma.FinalSize1}
		The final state of the epidemic model satisfies, for each $i=1,2$,
			\begin{enumerate}[label=\roman*)]
				\item\label{item.i} $S^\infty_i = \Nt_i-R^\infty_i - D^\infty_i$
				\item\label{item.ii} $R^\infty_i =\frac{\gamma_i}{\gamma_i+\mu_i}(S_{i0}+I_{i0} -S^\infty_i)+R_{i0}$
				\item\label{item.iii} $D^\infty_i =\frac{\mu_i}{\gamma_i+\mu_i}(S_{i0}+I_{i0} -S^\infty_i)+D_{i0}$
			\end{enumerate}
	\end{lemma}

	\begin{proof}
		The item \ref{item.i} follows from Theorem \ref{thm:extinction} and the definition of $\Nt_i$.
		
		For the item \ref{item.ii}, we can use the equations for $R$ and $D$ in \eqref{Eq.SIRDm} and obtain
		\[	R_i'(t)+D_i'(t)=(\gamma_i+\mu_i)I_i(t) = \frac{\gamma_i+\mu_i}{\gamma_i}\gamma_iI_i(t) = \frac{\gamma_i+\mu_i}{\gamma_i}R'_i(t),\]
		which implies that
		\[R_i(t)+D_i(t)-R_{i0}-D_{i0}=\frac{\gamma_i+\mu_i}{\gamma_i}(R_i(t)-R_{i0}), \qquad \forall \; t>0.\]
		Therefore, as $t\to+\infty$ we conclude that
		\[R_i^\infty+D_i^\infty-R_{i0}-D_{i0}=\frac{\gamma_i+\mu_i}{\gamma_i}(R_i^\infty-R_{i0}).\]
		Using item \ref{item.i} and noting that $\Nt_i-R_{i0}-D_{i0}=S_{i0}+I_{i0}$, we can write
		\[S_{i0}+I_{i0}-S_i^\infty=\frac{\gamma_i+\mu_i}{\gamma_i}(R_i^\infty-R_{i0})\]
		and thus
		\[R_i^\infty=\frac{\gamma_i}{\gamma_i+\mu_i}(S_{i0}+I_{i0}-S_i^\infty)+R_{i0}.\]
		Finally, item \ref{item.iii} follows from items \ref{item.i} and \ref{item.ii}.
	\end{proof}

	For each $i=1,2$, in order to obtain more information about the the asymptotic values $S_i^\infty$, let us begin with the equation for $S_i(t)$ in \eqref{Eq.SIRDm}.  Since $S_1(0),S_2(0)>0$, by continuity we can take $\ln S_i(t)$ at least for $t$ sufficiently small. Thus, we have that
	\begin{equation}\label{Eq.logS}
		\frac{d}{dt}(\ln S_i(t)) = -\left(\beta_{i1}\frac{I_1(t)}{N_1(t)} +\beta_{i2}\frac{I_2(t)}{N_2(t)} \right)
	\end{equation}  
	By adding the first three equations we obtain
	\[\frac{d}{dt}N_j(t)=  \frac{d}{dt}(S_j(t)+I_j(t)+R_j(t)) = -\mu_j I_j(t) , \]
	for each $j=1,2$, which implies
	\begin{equation}\label{Eq.logN}
		\frac{d}{dt}(\ln N_j(t))= -\mu_j\frac{I_j(t)}{N_j(t)}.
	\end{equation}
	The equation above is valid as long as there is an individual alive in each group. Therefore, by \eqref{Eq.logS} and \eqref{Eq.logN}
	\begin{equation}\label{Eq.lnConstante}
		\frac{d}{dt}\Big(\ln S_i(t) -\frac{\beta_{i1}}{\mu_1}\ln N_1(t) -\frac{\beta_{i2}}{\mu_2}\ln N_2(t)\Big) = 0 .
	\end{equation}

\begin{theorem}\label{Thm.PositivityS}
Assume \ref{H1} and $S_{0}\gg0$. Let $(S,I,R,D)$ be the solution of \eqref{Eq.SIRDm} with initial data in $\mathcal D$. Then, for each $i=1,2$:
\begin{enumerate}[label=\roman*)]
\item $S_i(t)>0$ and $N_i(t)>0$ for all $t\ge 0$;
\item \label{eq.F.constant} the function $\F(t)=(F_1(t),F_2(t))$ defined by
\[
F_i(t):=\frac{S_i(t)}{N_1(t)^{\frac{\beta_{i1}}{\mu_1}}\,N_2(t)^{\frac{\beta_{i2}}{\mu_2}}}
\]
is well defined on $[0,\infty)$ and constant in time, with
\[
F_i(t)\equiv F_i(0)
= \frac{S_{i0}}{N_1(0)^{\frac{\beta_{i1}}{\mu_1}}\,N_2(0)^{\frac{\beta_{i2}}{\mu_2}}}>0;
\]
\item \label{eq.Sinf-positive} in particular, we have
\[
S_i^\infty
= F_i(0)\,\bigl(N_1^\infty\bigr)^{\frac{\beta_{i1}}{\mu_1}}
              \bigl(N_2^\infty\bigr)^{\frac{\beta_{i2}}{\mu_2}}>0.
\]
\end{enumerate}
\end{theorem}

\begin{proof}
From the $S_i$-equation in \eqref{Eq.SIRDm} we have
\[
S_i'(t)
= -\left(\beta_{i1}\frac{I_1(t)}{N_1(t)} + \beta_{i2}\frac{I_2(t)}{N_2(t)} \right) S_i(t).
\]
As long as $N_1(t)$ and $N_2(t)$ are defined, this yields
\[
S_i(t)
= S_{i0}\exp\!\left(
  -\int_0^t\left(\beta_{i1}\frac{I_1(\tau)}{N_1(\tau)}
               +\beta_{i2}\frac{I_2(\tau)}{N_2(\tau)}\right)\,d\tau
\right),
\]
so $S_i(t)>0$ for all $t\ge 0$. Consequently
\[
N_i(t)=S_i(t)+I_i(t)+R_i(t)\ge S_i(t)>0,\qquad t\ge 0,
\]
which proves (i) and shows that $\ln S_i(t)$ and $\ln N_i(t)$ are well defined for all $t\ge 0$.

Therefore, from \eqref{Eq.lnConstante} the quantity inside the parentheses is constant in time for all $t\ge 0$, and exponentiating gives
\[
 \frac{S_i(t)}{N_1(t)^{\frac{\beta_{i1}}{\mu_1}}\,N_2(t)^{\frac{\beta_{i2}}{\mu_2}}}
=F_i(t)
= F_i(0)
= \frac{S_{i0}}{N_1(0)^{\frac{\beta_{i1}}{\mu_1}}\,N_2(0)^{\frac{\beta_{i2}}{\mu_2}}}>0,
\]
which proves (ii).

Finally, by Theorem \eqref{thm:extinction}, we can take $t\to\infty$ in the identity defining $F_i(t)$ and using its constancy to obtain
\[
S_i^\infty
= F_i(0)\,\bigl(N_1^\infty\bigr)^{\frac{\beta_{i1}}{\mu_1}}
              \bigl(N_2^\infty\bigr)^{\frac{\beta_{i2}}{\mu_2}},
\]
which is strictly positive because $F_i(0)>0$ and $N_j^\infty=S^\infty_j+R^\infty_j\geq R_j^\infty> 0$. This establishes (iii).
\end{proof}

From item \ref{eq.Sinf-positive} from Theorem \ref{Thm.PositivityS}, we use that $N_j^\infty= S_j^\infty+R_j^\infty$, for $j=1,2$, to write
	\begin{equation}\label{Eq.Si.infty1}
		S_i^\infty=F_i(0)(S_1^\infty+R_1^\infty)^{\frac{\beta_{i1}}{\mu_1}}(S_2^\infty+R_2^\infty)^{\frac{\beta_{i2}}{\mu_2}}, \qquad i=1,2.
	\end{equation}
	By item \ref{item.ii} of Lemma \ref{Lemma.FinalSize1} and $S_{j0}+I_{j0}=N_j(0)-R_{j0} $, we know that for each $j=1,2$
	\begin{equation}\label{Eq.Rj.infty}
		R_j^\infty=(1-\epsilon_j)(-S_j^\infty+N_j(0)-R_{j0} )+R_{j0}, \quad\mbox{ where }\quad \epsilon_j=\frac{\mu_j}{\gamma_j+\mu_j}
	\end{equation}
	To simplify the notation, let us define
	\begin{equation}\label{Eq.Wj}
		Z_j=(1-\epsilon_j)N_j(0)+\epsilon_jR_{j0}, \qquad j=1,2.
	\end{equation}
	Thus, from \eqref{Eq.Si.infty1}, \eqref{Eq.Rj.infty} and item \ref{eq.F.constant} from Theorem \ref{Thm.PositivityS} we obtain that for each $j=1,2$
	\begin{equation}\label{Eq.Sinfty2}
		S_i^\infty=S_{i0}\left(\frac{\epsilon_1S_1^\infty+Z_1}{N_1(0)}\right)^{\frac{\beta_{i1}}{\mu_1}}\left(\frac{\epsilon_2S_2^\infty+Z_2}{N_2(0)}\right)^{\frac{\beta_{i2}}{\mu_2}}, \quad i=1,2.
	\end{equation}

which is a closed-form expression linking the final size to the initial conditions parameters of the model.

	\begin{definition}
	    Let $T:\R_+^2\to \R_+^2$ with $T(x_1,x_2)=(T_1(x_1,x_2),T_2(x_1,x_2))$ be the map given by
    \begin{equation}\label{def.T}
		\left\{ \begin{array}{l}
		     T_1(x_1,x_2) = S_{10}\left(\frac{\epsilon_1x_1+Z_1}{N_1(0)}\right)^{\frac{\beta_{11}}{\mu_1}}\left(\frac{\epsilon_2x_2+Z_2}{N_2(0)}\right)^{\frac{\beta_{12}}{\mu_2}}  \\
		     T_2(x_1,x_2) = S_{20}\left(\frac{\epsilon_1x_1+Z_1}{N_1(0)}\right)^{\frac{\beta_{21}}{\mu_1}}\left(\frac{\epsilon_2x_2+Z_2}{N_2(0)}\right)^{\frac{\beta_{22}}{\mu_2}}
		\end{array}		\right. ,
	\end{equation}
	where $\epsilon_j$ and $Z_j$ are defined above in \eqref{Eq.Rj.infty} and \eqref{Eq.Wj},
	\end{definition}

    Therefore, we obtained the following result:

    \begin{theorem}\label{thm:general-final}
Assume \ref{H1} and $S_0\gg0$, $I_0\ge0$. Then the final susceptible state $S^\infty=(S_1^\infty,S_2^\infty)$ is positive and solves the algebraic system
\[
S^\infty=T(S^\infty)
\]
i.e., $S^\infty$ is a fixed point of $T$ in $[0,S_{10}]\times[0,S_{20}]$.

\end{theorem}

	\begin{theorem}\label{Thm.mapT}
		Let $T:\R_+^2\to\R_+^2$ be the map defined in \eqref{def.T} and $I_0>0$. Then
		\begin{enumerate}[label=\roman*)]
			\item\label{Thm.i} $X\ll Y \Rightarrow T(X)\ll T(Y)$.  Also  $0\ll T(0) \ll T(S_0)\ll S_0$.
			\item\label{Thm.ii} There exist $0\ll S^-\leq S^+ \ll S_0$ such that $T(S^\pm)=S^\pm$.
			\item\label{Thm.iii} If \ref{H2a} holds, the Jacobian of $T$ is componentwise increasing, i.e, 
            \[X\le Y \Rightarrow \frac{\partial T_i}{\partial x_j}(X)\le \frac{\partial T_i}{\partial x_j}(Y)\;, \;\; \mbox{ for all }i,j.\]
            Furthermore, for the fixed points $S^\pm$, there holds $JT(S^\pm)=\diag(S^\pm) \B \M^{-1} \diag(\E)$, where
            $\E=\left(\epsilon_1(\epsilon_1S^\pm_1+Z_1)^{-1},\epsilon_2(\epsilon_2S^\pm_2+Z_2)^{-1}\right)$
			
		\end{enumerate}

	\end{theorem}

	\begin{proof}
		The first and second inequalities of item \ref{Thm.i} follows from the definition of $T$. For the last one, note that for each $i=1,2$, $T_i(S_0)$ can be rewritten 
		\begin{align*}
			T_i(S_{10},S_{20})&= S_{i0}\prod_{j=1}^{2}\left(\frac{\epsilon_jS_{j0}+Z_j}{N_j(0)}\right)^{\frac{\beta_{ij}}{\mu_j}}\\
			&= S_{i0}\prod_{j=1}^{2}\left(\frac{\epsilon_j(S_{j0}+R_{j0})+(1-\epsilon_j)N_j(0)}{N_j(0)}\right)^{\frac{\beta_{ij}}{\mu_j}}\\
			&= S_{i0}\prod_{j=1}^{2}\left(\epsilon_j\left(1-\frac{I_{j0}}{N_j(0)}\right)+1-\epsilon_j\right)^{\frac{\beta_{ij}}{\mu_j}}\\
			&= S_{i0}\prod_{j=1}^{2}\left(1-\epsilon_j\frac{I_{j0}}{N_j(0)}\right)^{\frac{\beta_{ij}}{\mu_j}}
		\end{align*}
		Therefore, since $\epsilon_j<1$ for $j=1,2$, we obtain that $T_i(S_{10},S_{20})= S_{i0}$ for $i=1$ or $i=2$ if and only if $I_{j0}=0$ for $j=1,2$. Therefore, under the assumption that $I_0>0$, then at least one of the terms $1-\epsilon_j\frac{I_{j0}}{N_j(0)}<1$.

		For item \ref{Thm.ii}, we use \ref{Thm.i} and note that, for every $n\in \mathbb{N}$
		\[    0\ll T(0) \ll \cdots \ll T^n(0)\ll \cdots \ll T^n(S_0) \ll \cdots  T(S_0)\ll S_0.  \]
		Therefore, there exist the limits
		\[ S^- = \lim_{n\to\infty} T^n(0) \;\;,\;\;\; S^+ = \lim_{n\to\infty} T^n(S_0), \]
		they satisfy $S^-\leq S^+$  and are fixed points of $T$ on $[0,S_{10}]\times[0,S_{20}] $, since $T$ is continuous. Note that, due to the monotonicity of $T$, there is no other fixed points of $T$ in $[0,S_{10}]\times[0,S_{20}] \backslash [S_1^-,S_1^+]\times[S_2^-,S_2^+]$.
		
		Finally, the item \ref{Thm.iii} also follows from the definition of $T$. 		
	\end{proof}

    \begin{corollary}[Bracketing of the final susceptible state]\label{cor:bracketing-final-size}
Assume \ref{H1} and $S_0\gg 0$, $I_0\ge 0$. Let $T:\mathbb{R}_+^2\to\mathbb{R}_+^2$ be the map defined in \eqref{def.T}, and let $S^-,S^+\in(0,S_0)$ be the fixed points given by Theorem~\ref{Thm.mapT}.
Then the final susceptible state $S^\infty=(S_1^\infty,S_2^\infty)$ of system \eqref{Eq.SIRDm} satisfies
\[
S^- \;\le\; S^\infty \;\le\; S^+.
\]
\end{corollary}

    The next result follows some arguments used for the SIR model in \cite{Magal2} and some new ones.

	\begin{theorem}[Final size]\label{Theo.FinalSize}
		Assume that \ref{H1} and any of the following \ref{H2a},\ref{H2b},\ref{H2c},\ref{H2d} hold. If $S_0\gg 0$ and $I_0>0$, then the final state of the epidemic model satisfies
		\[0\neq S^\infty = \lim_{n\to \infty} T^n(0)\]
		
	\end{theorem}

	\begin{proof}
		WWe will show that, under our hypotheses, $T$ admits a unique fixed point. As a consequence, this fixed point can be obtained by iterating $T$ starting from any initial point, in particular from $0$.
		
		Under the hypotheses \ref{H1} and \ref{H2a}, let us suppose by contradiction that $S^-<S^+$. 
By Taylor's theorem we can write
\begin{align}
  S^+ - S^- 
  &= T(S^+) - T(S^-) \nonumber \\
  &= \int_0^1 JT\bigl(S^-+\tau(S^+ - S^-)\bigr)\,(S^+ - S^-)\,d\tau. \label{Eq.S+S--H2a}
\end{align}
Since $S^- \le S^-+\tau(S^+ - S^-)\le S^+$ for all $\tau\in[0,1]$ and, under \ref{H2a}, the Jacobian
$JT$ is componentwise increasing (Theorem~\ref{Thm.mapT}(iv)), we have
\[
  JT\bigl(S^-+\tau(S^+ - S^-)\bigr) \le JT(S^+)
\]
componentwise for all $\tau\in[0,1]$. Because $S^+ - S^-\gg0$ and all entries of $JT$ are nonnegative,
\eqref{Eq.S+S--H2a} implies
\begin{equation}\label{Eq.S+S--ineq}
  S^+ - S^- \le JT(S^+)(S^+ - S^-).
\end{equation}
By \ref{H1} and Theorem~\ref{Thm.mapT}(iv), the matrix $JT(S^+)$ is positive. Hence, by the
Perron--Frobenius theorem, it has a dominant eigenvalue $\lambda>0$ with a strictly positive left
eigenvector $W^\top\gg0$:
\[
  W^\top JT(S^+) = \lambda W^\top.
\]
Multiplying \eqref{Eq.S+S--ineq} on the left by $W^\top$ we obtain
\[
  W^\top(S^+ - S^-) \le W^\top JT(S^+)(S^+ - S^-)
                    = \lambda\,W^\top(S^+ - S^-).
\]
Since $S^+ - S^-\gg0$ and $W\gg0$, we have $W^\top(S^+ - S^-)>0$, and therefore
\begin{equation}\label{Eq.lambda-ge-1}
  \lambda \ge 1.
\end{equation}

Next, using the fact that $S^+\le S_0$, we write
\begin{align}
  T(S_0) - S^+
  &= T(S_0) - T(S^+) \nonumber \\
  &= \int_0^1 JT\bigl(S^+ + \tau(S_0 - S^+)\bigr)\,(S_0 - S^+)\,d\tau. \label{Eq.TS0Sp-int}
\end{align}
Now $S^+ \le S^+ + \tau(S_0 - S^+) \le S_0$ for all $\tau\in[0,1]$, and again by the componentwise
monotonicity of $JT$ we have
\[
  JT(S^+) \le JT\bigl(S^+ + \tau(S_0 - S^+)\bigr)
\]
for all $\tau\in[0,1]$. Since $S_0 - S^+\ge 0$ and $JT$ has nonnegative entries, \eqref{Eq.TS0Sp-int}
yields
\begin{equation}\label{Eq.TS0Sp-ineq}
  T(S_0) - S^+ 
  \ge JT(S^+)(S_0 - S^+).
\end{equation}
Multiplying \eqref{Eq.TS0Sp-ineq} on the left by $W^\top$ and using $W^\top JT(S^+)=\lambda W^\top$,
we find
\[
  W^\top(T(S_0) - S^+) \ge \lambda\,W^\top(S_0 - S^+).
\]
Subtracting $W^\top(S_0 - S^+)$ from both sides and using \eqref{Eq.lambda-ge-1}, we obtain
\begin{equation}\label{Eq.WTS0S0}
  W^\top\bigl(T(S_0) - S_0\bigr)
  = W^\top(T(S_0) - S^+) - W^\top(S_0 - S^+)
  \ge (\lambda - 1)\,W^\top(S_0 - S^+) \ge 0.
\end{equation}

On the other hand, item~\ref{Thm.i} of Theorem~\ref{Thm.mapT} tells us that, when $I_0>0$,
we have $T(S_0)\ll S_0$. Since $W\gg0$, this implies
\[
  W^\top\bigl(T(S_0) - S_0\bigr) < 0,
\]
which contradicts \eqref{Eq.WTS0S0}. Therefore our assumption $S^-<S^+$ was false, and we must
have $S^- = S^+$. In particular, under \ref{H1} and \ref{H2a} the fixed point of $T$ in $(0,S_0)$
is unique.
		
Now, assuming \ref{H2b}  let us prove that $S^+\leq S^-$. We begin by showing that $T$ is sub-$\delta$-homogeneous for $\delta=\max\left\{\frac{\beta_{11}}{\mu_1}+\frac{\beta_{12}}{\mu_2}, \frac{\beta_{21}}{\mu_1}+\frac{\beta_{22}}{\mu_2}\right\}$, that is, for every $\eta\in(0,1)$ and $X\in \R^2_+$, there holds $\eta^\delta T(X)\ll T(\eta X)$.

For this, let $\eta\in(0,1)$ and $i\in\{1,2\}$. Then for any $X=(x_1,x_2)\in\R^2_+$
\begin{eqnarray*}
\eta^\delta T_i(X)&=&\eta^\delta T_i(x_1,x_2)\\
&\leq&\eta^{\frac{\beta_{i1}}{\mu_1}+\frac{\beta_{i2}}{\mu_2}} S_{i0}\left(\frac{\epsilon_1x_1+Z_1}{N_1(0)}\right)^{\frac{\beta_{i1}}{\mu_1}}\left(\frac{\epsilon_2x_2+Z_2}{N_2(0)}\right)^{\frac{\beta_{i2}}{\mu_2}}\\
&\leq& S_{i0}\left(\frac{\epsilon_1\eta x_1+\eta Z_1}{N_1(0)}\right)^{\frac{\beta_{i1}}{\mu_1}}\left(\frac{\epsilon_2\eta x_2+\eta Z_2}{N_2(0)}\right)^{\frac{\beta_{i2}}{\mu_2}}\\
&<&S_{i0}\left(\frac{\epsilon_1\eta x_1+ Z_1}{N_1(0)}\right)^{\frac{\beta_{i1}}{\mu_1}}\left(\frac{\epsilon_2\eta x_2+Z_2}{N_2(0)}\right)^{\frac{\beta_{i2}}{\mu_2}}\\
&=&T_i(\eta X)
\end{eqnarray*}	
and the sub-$\delta$-homogeneity is proved.

Let $\eta_*=\sup\{\eta>0\;|\; \eta S^+\leq S^-\}$. The supremum is well defined since $0\ll S^-\leq S^+$. Suppose, by contradiction, that $\eta_*<1$. Therefore, by the sub-$\delta$-homogeneity and monotonicity of $T$ we obtain
\[\eta_*^\delta S^+=\eta_*^\delta T(S^+)\ll T(\eta_* S^+)\leq T(S^-)=S^-\]
which contradicts the fact that $\eta_*$ is the supremum since $\eta_*^\delta>\eta_*$ when \ref{H2b} holds. Therefore, we conclude that $\eta_*\geq 1$, and thus $S^+\leq \eta_*S^+\leq S^-$. Since we already know that the oposite inequality holds, we obtain that $S^-=S^+$.



When \ref{H2c} holds, we define $H: (0,S_{10}]\times(0,S_{20}] \to\mathbb{R}^2$ for $X=(x_1,x_2)$ by
\[
 H_i(X)=\log x_i-\log T_i(X),\;\; i=1,2 
\]
Note that fixed points of $T$ are zeros of $H$. By Theorem \ref{Thm.mapT}, we already know that there exist $S^\pm$ such that $H(S^\pm)=0$.

For any $i,j\in\{1,2\}$, we have
\begin{equation*}
\frac{\partial H_i}{\partial x_j}(X)=
\frac{\delta_{ij}}{x_i}-\frac{\beta_{ij}}{\mu_j}\cdot\frac{\varepsilon_j}{\varepsilon_j x_j+Z_j}, \; 
\end{equation*}
where $\delta_{ij}$ is the Kronecker delta. We can see that when $i\neq j$, $\frac{\partial H_i}{\partial x_j}(X)\leq 0$. For the other terms,
\begin{eqnarray*}
    \frac{\partial H_i}{\partial x_i}(X) &=& \frac{1}{x_i}-\frac{\beta_{ii}}{\mu_i}\cdot\frac{\varepsilon_i}{\varepsilon_i x_i+Z_i} \\
    &\geq & \frac{1}{S_{i0}}-\frac{\beta_{ii}}{\mu_i}\cdot\frac{\varepsilon_i}{Z_i}\\
    &> & \frac{\beta_{ij}}{\mu_j}\cdot\frac{\varepsilon_j}{Z_j}\\
    &\geq& \frac{\beta_{ij}}{\mu_j}\cdot\frac{\varepsilon_j}{\varepsilon_j x_j+Z_j}\\
    &=& \left|\frac{\partial H_i}{\partial x_j}(X)\right|
\end{eqnarray*}
for $j\neq i$, where the strict inequality comes from the definition of $\varepsilon_i$ and the hypothesis \ref{H2c}. Therefore, we obtain that the Jacobian matrix $JH(X)$ has positive diagonal terms and is strictly diagonally dominant on $(0,S_1(0)]\times(0,S_2(0)]$. This dominance is actually uniform since the strict inequality does not depend on $X$. Also, the inequality above implies
\[\det JH = \frac{\partial H_1}{\partial x_1}\frac{\partial H_2}{\partial x_2}-\frac{\partial H_2}{\partial x_1}\frac{\partial H_1}{\partial x_2} >\left|\frac{\partial H_1}{\partial x_2}\right|\left|\frac{\partial H_2}{\partial x_1}\right|-\frac{\partial H_2}{\partial x_1}\frac{\partial H_1}{\partial x_2} \ge 0. \]
Therefore, all principal minors of $JH(X)$ are positive on the rectangular region of definition of $H$. By the Gale-Nikaidô Theorem (\cite{GaleNikaido}), it follows that $H$ is injective, and therefore it assumes zero at only one point.

Finally, for the case where \ref{H2d} holds, we will consider the change of variables $U=\log X$,i.e., $(u_1,u_2)=(\log x_1, \log x_2)$,   $\Omega = (-\infty, \log S_{10}]\times (-\infty, \log S_{10}]$ and define the function $G:\Omega\to\Omega$ by $G_i(U)=\log T_i(e^{u_1},e^{u_2})$. This function is well defined and has the property that $U$ is a fixed point for $G$ if and only if $e^U$ is a fixed point for $T$.
By the definition we have for any $i,j$:
\[\frac{\partial G_i}{\partial u_j}(U) = \frac{\beta_{ij}\varepsilon_j e^{u_j}}{\mu_j(\varepsilon_j e^{u_j}+Z_j)} \leq \frac{\beta_{ij}\varepsilon_j S_{j0}}{\mu_jZ_j}.\]
Thus, the hypothesis implies that the $L^\infty$ norm of the Jacobian of $G$ can be estimated 
\[\|JG(U)\|_\infty \leq \max_{i=1,2}\sum_{j=1}^2 \frac{\beta_{ij}}{\mu_j}\cdot\frac{\varepsilon_j S_{j0}}{Z_j}
<1.\]
Therefore, by the Mean Value Theorem the function $G$ is Lipschitz with constant less than 1, which implies that $G$ is a contraction on $\Omega$. By the Banach Fixed Point Theorem, it has only one fixed point.
\end{proof}

\begin{remark}
When $R_{10}=R_{20}=0$, the hypotheses \ref{H2c} and \ref{H2d} can be rewritten in a simpler way:
\[\mbox{\ref{H2c}} \;\;\frac{\beta_{i1}}{\gamma_1N_1(0)}+\frac{\beta_{i2}}{\gamma_2N_2(0)}<\frac{1}{S_{i0}}\;,\;\; i=1,2;\]
\[\mbox{\ref{H2d}} \;\;\frac{\beta_{i1}}{\gamma_1}\frac{S_{10}}{N_1(0)}+\frac{\beta_{i2}}{\gamma_2}\frac{S_{20}}{N_2(0)}<1\;,\;\; i=1,2;\]
\end{remark}

\begin{remark}
The final size $S^\infty$ can be approximated by iterating $T$ from any initial point $X\in[0,S_0]$.
\end{remark}	

\begin{remark}\label{rem:open-uniqueness}
The uniqueness of the fixed point of the map $T$ in Theorem~\ref{Theo.FinalSize}
is currently proved only under the structural assumptions \ref{H2a}-\ref{H2d}.
Outside these regimes, we do not know whether $T$ has a unique fixed point in
the biologically relevant rectangle $[0,S_0]$. We have also not been able to construct explicit examples of
the map $T$ arising from the SIRD model \eqref{Eq.SIRDm} (that is,
$\varepsilon_j$ and $Z_j$ linked to the epidemiological parameters and initial
conditions as in \eqref{Eq.Rj.infty} and\eqref{Eq.Wj}) that admit more than one fixed
point in $(0,S_0)$. From an epidemiological perspective, it is also reasonable to expect that a
large class of acute infections of interest will satisfy hypothesis \ref{H2a},
at least approximately, since $\beta_{ij}$ measures the typical rate of
transmission whereas $\mu_j$ measures the rate of disease-induced mortality.
\end{remark}

As an elementary application of the previous result, we prove that $S^\infty$ depends monotonically on the transmission rates $\beta_{ij}$.

\begin{proposition}\label{prop:monotonicity-order}
Consider two sets of
transmission parameters
\[
\beta=(\beta_{ij})_{i,j=1,2},\qquad \tilde\beta=(\tilde\beta_{ij})_{i,j=1,2},
\]
with $\beta_{ij},\tilde\beta_{ij}\ge0$. Let $T^\beta$ denote the map $T$
defined in \eqref{def.T} using the coefficients $\beta_{ij}$, and $T^{\tilde\beta}$
the analogous map defined with the coefficients $\tilde\beta_{ij}$. 

Assume that $T^\beta$ and $T^{\tilde\beta}$ have unique fixed points, denoted by $S^\infty(\beta), S^\infty(\tilde\beta) \in (0,S_0)$ respectively (this is the case when $\beta$ and $\tilde\beta$ are under the hypothesis of Theorem~\ref{Theo.FinalSize}).

If
\[
\tilde\beta_{ij} \;\ge\; \beta_{ij}\quad\text{for all }i,j\in\{1,2\},
\]
then
\[
S^\infty(\tilde\beta) \;\le\; S^\infty(\beta).
\]
In particular, the final susceptible state is \emph{nonincreasing}
in each parameter $\beta_{ij}$: increasing any transmission coefficient can only
decrease the final number of susceptibles in each group.
\end{proposition}

\begin{proof}
Fix $\beta$ and $\tilde\beta$ as in the statement. We make the dependence of
$T$ on the parameters explicit by using the definition of $T$:
\[
T_i^\beta(x_1,x_2)
= S_{i0}\prod_{j=1}^2\left(
\frac{\varepsilon_j x_j + Z_j}{N_j(0)}
\right)^{\frac{\beta_{ij}}{\mu_j}},
\qquad i=1,2.
\]
For any $X=(x_1,x_2)\in[0,S_0]$ and each $j$, we have
\[
0 < \frac{\varepsilon_j x_j + Z_j}{N_j(0)} \le 1.
\]
Indeed, by construction $Z_j=(1-\varepsilon_j)N_j(0)+\varepsilon_j R_{j0}\le N_j(0)$
and $0\le x_j\le S_{j0}\le N_j(0)$, so
\[
\varepsilon_j x_j + Z_j
= \varepsilon_j x_j + (1-\varepsilon_j)N_j(0)+\varepsilon_j R_{j0}
\le \varepsilon_j N_j(0)+(1-\varepsilon_j)N_j(0) = N_j(0).
\]

Therefore, if $\tilde\beta_{ij}\ge\beta_{ij}$, then
\[
\left(
\frac{\varepsilon_j x_j + Z_j}{N_j(0)}
\right)^{\frac{\tilde\beta_{ij}}{\mu_j}}
\;\le\;
\left(
\frac{\varepsilon_j x_j + Z_j}{N_j(0)}
\right)^{\frac{\beta_{ij}}{\mu_j}},
\]
with strict inequality whenever $\tilde\beta_{ij}>\beta_{ij}$ and the
corresponding factor is strictly less than $1$. Multiplying over $j=1,2$, we
obtain, for each $i$ and all $X\in[0,S_0]$,
\[
T_i^{\tilde\beta}(X)
= S_{i0}\prod_{j=1}^2\left(
\frac{\varepsilon_j x_j + Z_j}{N_j(0)}
\right)^{\frac{\tilde\beta_{ij}}{\mu_j}}
\;\le\;
S_{i0}\prod_{j=1}^2\left(
\frac{\varepsilon_j x_j + Z_j}{N_j(0)}
\right)^{\frac{\beta_{ij}}{\mu_j}}
= T_i^{\beta}(X).
\]
In vector notation this means
\begin{equation}\label{eq:Tbeta-comp}
T^{\tilde\beta}(X) \;\le\; T^{\beta}(X)
\qquad\text{for all }X\in[0,S_0].
\end{equation}

Now consider the iterates starting from $0$:
\[
X_{n+1}^\beta = T^\beta(X_n^\beta),\quad X_0^\beta=0,
\qquad
X_{n+1}^{\tilde\beta} = T^{\tilde\beta}(X_n^{\tilde\beta}),\quad X_0^{\tilde\beta}=0.
\]
By the monotonicity of $T^\beta$ in $X$ (Theorem~\ref{Thm.mapT}) and the
comparison \eqref{eq:Tbeta-comp}, we obtain by induction on $n$ that
\[
X_n^{\tilde\beta} \;\le\; X_n^\beta\qquad\text{for all }n\ge0.
\]
Indeed, for $n=0$ we have $X_0^{\tilde\beta}=X_0^\beta=0$. Suppose
$X_n^{\tilde\beta}\le X_n^\beta$; then, using the monotonicity of
$T^{\tilde\beta}$ in $X$ and \eqref{eq:Tbeta-comp},
\[
X_{n+1}^{\tilde\beta}
= T^{\tilde\beta}(X_n^{\tilde\beta})
\le T^{\tilde\beta}(X_n^\beta)
\le T^{\beta}(X_n^\beta)
= X_{n+1}^\beta.
\]

By Theorem~\ref{Thm.mapT}, the sequences $X_n^\beta$ and $X_n^{\tilde\beta}$
converge, respectively, to the respective fixed points
\[
S^\infty(\beta) = \lim_{n\to\infty}X_n^\beta,
\qquad
S^\infty(\tilde\beta) = \lim_{n\to\infty}X_n^{\tilde\beta}.
\]
Passing to the limit as $n\to\infty$ in the inequality
$X_n^{\tilde\beta}\le X_n^\beta$ yields
\[
S^\infty(\tilde\beta) \;\le\; S^\infty(\beta).
\]

\end{proof}

\begin{remark}
    Although the assumptions H2a, H2b, and H2c are sufficient to establish the uniqueness of the fixed point, they do not by themselves yield any quantitative estimate on the speed of convergence of the iterates $T^n(0)$. In particular, these conditions ensure injectivity or diagonal dominance properties of the associated Jacobian, but they do not imply that $T$ (or its logarithmic transform) is a contraction. As a consequence, no uniform rate of convergence can be deduced from them, and the iterates may in principle converge arbitrarily slowly. The only assumption that directly provides a convergence rate is H2d, which guarantees that $T$ is a contraction in logarithmic coordinates and thus yields a global linear rate of convergence for the sequence $T^n(0)$. For this last case, we can write some standard numerical analysis estimates as in the proposition below.
\end{remark}

\begin{proposition}
Assume H2d holds, i.e.
\[
\Lambda \;:=\; \max_{i=1,2}\sum_{j=1}^2\frac{\beta_{ij}}{\gamma_j+\mu_j}\,\frac{S_{j0}}{Z_j}\;<\;1.
\]
Let $X_{n+1}=T(X_n)$ with $X_0=0$ and define $U_n:=\log X_n$ for $n\ge1$ (so $X_1=T(0)\gg0$). Let $S^\infty$ be the unique fixed point of $T$ in $\Omega$, and put $U^\infty:=\log S^\infty$.

\begin{enumerate}[label=(\roman*)]
\item (rate with $S^\infty$)
For all $n\ge 1$, let $\delta_n=\Lambda^{\,n-1}\,\|U_{1}-U^\infty\|_\infty$. Then we have
\[
\|U_{n}-U^\infty\|_\infty \;\le\; \delta_n.
\]
In particular, we can estimate each coordinate in the original variables
\[
|x_{n,i}-S_i^\infty|\ \le\ \bigl(e^{\delta_n}-1\bigr)\,x_{n,i}.
\]

\item (without $S^\infty$)
For all $n\ge 2$, let $\tau_n = \frac{\Lambda}{1-\Lambda}\,\|U_n-U_{n-1}\|_\infty$. Then
\[
\|U_{n}-U^\infty\|_\infty \;\le\; \tau_n.
\]
Also, componentwise in the $X$ variables

\[
|x_{n,i}-S_i^\infty| \;\le\; \bigl(e^{\tau_n}-1\bigr)\,x_{i,n}. 
\]
\end{enumerate}
\end{proposition}
\begin{proof}
By the proof of Theorem \ref{Theo.FinalSize}, we know that \ref{H2d} implies that $G(U):=\log T(e^U)$ is a $\Lambda$-contraction on the rectangle $\Omega$ with respect to $\|\cdot\|_\infty$. Since $X_1=T(0)\gg 0$, all iterates $U_n=G(U_{n-1})$ are well-defined for $n\ge 2$ and $U^\infty$ solves $U^\infty=G(U^\infty)$. 

\begin{enumerate}[label=(\roman*)]
\item Since $G$ is a $\Lambda$ contraction,
\[\|U_{n}-U^\infty\|_\infty = \|G(U_{n-1})-G(U^\infty)\|_\infty \le \Lambda \|U_{n-1}-U^\infty\|_\infty\le \cdots \le \Lambda^{n-1} \|U_{1}-U^\infty\|_\infty=\delta_n\]
which gives the first inequality. For each coordinate, this means that
\[0<\log \frac{S_i^\infty}{x_{n,i}} = |\log x_{n,i}-\log S_i^\infty| \leq \delta_n,\]
which gives 
\[1<\frac{S_i^\infty}{x_{n,i}}  \leq e^{\delta_n}.\]
This last one is equivalent to the desired inequality.

\item For any given $n$ and $k\ge 0$ we have $\|U_{n+k+1}-U_{n+k}\|_\infty = \Lambda^k\|U_{n+1}-U_{n}\|_\infty$. Using that
$U^\infty-U_n = \sum_{k=0}^{\infty}(U_{n+k+1}-U_{n+k})$, we obtain
\begin{eqnarray*}
\|U^\infty-U_n\|_\infty \;&\le&\; \sum_{k=0}^{\infty}\|U_{n+k+1}-U_{n+k}\|_\infty \;\le\; \left(\sum_{k=0}^{\infty}\Lambda^k\right)\|U_{n+1}-U_{n}\|_\infty\\
\;&\le&\; \frac{1}{1-\Lambda}\,\|U_{n+1}-U_n\|_\infty
\;\le\; \frac{\Lambda}{1-\Lambda}\,\|U_n-U_{n-1}\|_\infty = \tau_n.    
\end{eqnarray*}
The componentwise bounds for $x_{n,i}$ follow as in item i) with $\delta_n$ replaced by $\tau_n$.
\end{enumerate}
\end{proof}

\section{Basic reproduction number and final size}

Another key quantity in epidemic modeling is the basic reproduction number, denoted $\Ro $, which measures the expected number of secondary cases produced by a typical infectious individual in a fully susceptible population. While $\Ro $ governs the early dynamics of an epidemic, the final size is a descriptor of its long-term impact. The classical final size relation derived by Kermack and McKendrick (\cite[p. 708]{KmcK}) expresses the proportion of susceptible individuals who remain uninfected at the end of an epidemic as a solution to a nonlinear equation involving $\Ro $ and also the initial number of susceptible individuals.

In this section we relate the basic reproduction number $\Ro$ of the system \eqref{Eq.SIRDm} to the final state $S^\infty$, thus relating the early phase of the epidemic to its final stage. We first recall the usual next generation matrix construction and then derive a logarithmic identity connecting $\Ro$ to the final susceptible state.

Let $\Gamma=\diag(\gamma_1,\gamma_2)$ and $\M=\diag(\mu_1,\mu_2)$. According to the next generation matrix approach in \cite{NGM}, the basic reproduction number $\Ro$ is defined as the spectral radius of a matrix $\K$ of the form $FV^{-1}$, where $F$ collects the new infection terms and $V$ the transition terms out of the infectious compartments. For system \eqref{Eq.SIRDm}, the $2\times2$ next generation matrix is
\[
  \K = \begin{pmatrix}
    \dfrac{\beta_{11}}{\gamma_1+\mu_1} & \dfrac{\beta_{12}}{\gamma_2+\mu_2}\dfrac{S_{10}}{S_{20}}\\[0.4em]
    \dfrac{\beta_{21}}{\gamma_1+\mu_1}\dfrac{S_{20}}{S_{10}} & 	\dfrac{\beta_{22}}{\gamma_2+\mu_2}
  \end{pmatrix},
\]
so that
\[
  \Ro = \rho(\K),
\]
the spectral radius of $\K$. This matrix can be rewritten as
\[
  \K = \diag(S_0)\,\B\,(\Gamma+\M)^{-1}\,\diag(S_0)^{-1},
\]
and hence $\K$ is similar to
\[
  \K_1 := \B\,(\Gamma+\M)^{-1}.
\]
In particular, $\K_1$ has the same eigenvalues as $\K$ and therefore also has spectral radius $\Ro$. By hypothesis \ref{H1}, the matrix $\K_1$ is positive, so by the Perron–Frobenius theorem there exists a strictly positive left eigenvector $V^\top=(V_1,V_2)$ such that
\[
  V^\top\K_1 = \Ro\,V^\top.
\]

We now relate $\Ro$ to the final size of the epidemic.

\begin{theorem}[Final size identity involving $\Ro$]\label{thm:R0-finalsize}
Assume \ref{H1} and let $V\in\mathbb{R}^2$ be a positive left eigenvector of $\K_1=\B(\Gamma+\M)^{-1}$ associated with the largest eigenvalue $\Ro$. Then the final susceptible state $S^\infty$ and basic reproduction number $\Ro$ are related through the initial conditions by
\begin{equation}\label{eq:R0-finalsize}
 V^\top\ln\left(\frac{S_0}{S^\infty}\right)
  = \mathcal{R}_0V^\top\,(\Gamma\M^{-1}+I)\,\,\ln\left(\frac{(\Gamma+\M)N_0}{(\Gamma+\M)N_0+\M(S^\infty -S_0-I_0)}\right) 
\end{equation}
where $I$ is the $2\times2$ identity matrix and all logarithms and quotients are taken componentwise.
\end{theorem}

\begin{proof}
    Writing \eqref{Eq.lnConstante} in vector form, we know that for every $t\ge0$ the quantity
\[
  \ln S(t) - \B\,\M^{-1}\,\ln N(t)
\]
is constant in time, where $S(t)=(S_1(t),S_2(t))^\top$ and $N(t)=(N_1(t),N_2(t))^\top$, and the logarithms are taken componentwise. Therefore, evaluating at $t=0$ and passing to the limit $t\to+\infty$ we obtain
\[
  \ln S_0 - \B\,\M^{-1}\,\ln N_0
  = \ln S^\infty - \B\,\M^{-1}\,\ln N^\infty,
\]
that is,
\[
  \ln\left(\frac{S_0}{S^\infty}\right)
  = \B\,\M^{-1}\,\ln\left(\frac{N_0}{N^\infty}\right),
\]
where all divisions are componentwise. Since $\B=\K_1(\Gamma+\M)$, we can rewrite this as
\begin{equation}\label{eq:finalsize-K1.1}
  \ln\left(\frac{S_0}{S^\infty}\right)
  = \K_1\,(\Gamma+\M)\,\M^{-1}\,\ln\left(\frac{N_0}{N^\infty}\right).
\end{equation}

Multiplying the last identity on the left by the positive left eigenvector $V^\top$ of $\K_1$ associated with $\Ro$, we obtain the following final size relation.
\begin{equation}\label{eq:finalsize-K1}
  V^\top\ln\left(\frac{S_0}{S^\infty}\right)
  = \mathcal{R}_0V^\top\,(\Gamma\M^{-1}+I)\,\ln\left(\frac{N_0}{N^\infty}\right).
\end{equation}
This is already a relation between the $\mathcal{R}_0$ and the final state, but we can make it explicit in terms of $S^\infty$ and the initial conditions:

Recall that $N_i^\infty=S_i^\infty+R_i^\infty$, so we can use item \ref{item.ii} from Lemma \ref{Lemma.FinalSize1}:
\begin{align*}
  N_i^\infty &= S_i^\infty + N_i(0)-\frac{\gamma_i}{\gamma_i+\mu_i}S_i^\infty -\frac{\mu_i}{\gamma_i+\mu_i}(S_{i0}+I_{i0})\\
  &= N_i(0) +\frac{\mu_i}{\gamma_i+\mu_i}(S_i^\infty-S_{i0}-I_{i0})
\end{align*}
Therefore 
\begin{equation}\label{eq.Ni/Ninfty}
    \frac{N_i(0)}{N_i^\infty} = \frac{(\gamma_i+\mu_i)N_i(0)}{(\gamma_i+\mu_i)N_i(0) +\mu_i(S_i^\infty-S_{i0}-I_{i0})}
\end{equation}
Replacing \eqref{eq.Ni/Ninfty} into \eqref{eq:finalsize-K1} we obtain
\[V^\top\ln\left(\frac{S_0}{S^\infty}\right)
  = \mathcal{R}_0V^\top\,(\Gamma\M^{-1}+I)\,\,\ln\left(\frac{(\Gamma+\M)N_0}{(\Gamma+\M)N_0+\M(S^\infty -S_0-I_0)}\right)\]

\end{proof}


\section{Simulations}

We would like to illustrate some differences between the classical two-group SIRD model and the model \eqref{Eq.SIRDm}, as well as to show some different dynamics that these models can cover that the one group SIRD model does not. We also present an illustration of Theorem \ref{Theo.FinalSize}.

The classical SIRD system for two groups can be written as 

\begin{equation}\label{Eq.SIRDc}
		\left\{ \begin{array}{ll}
			S_1'  = -\left(\beta_{11}\frac{I_1}{\Nt_1} +\beta_{12}\frac{I_2}{\Nt_2} \right) S_1    ,& S_2'  = -\left(\beta_{21}\frac{I_1}{\Nt_1} +\beta_{22}\frac{I_2}{\Nt_2} \right) S_2 \\
			I_1'  =\left(\beta_{11}\frac{I_1}{\Nt_1} +\beta_{12}\frac{I_2}{\Nt_2} \right) S_1 - (\gamma_1+\mu_1)I_1  ,& I_2'  =\left(\beta_{21}\frac{I_1}{\Nt_1} +\beta_{22}\frac{I_2}{\Nt_2} \right) S_2  - (\gamma_2+\mu_2)I_2 \\
			R_1'  = \gamma_1 I_1 ,&R_2'  = \gamma_2 I_2 \\
			D_1'  = \mu_1 I_1 ,&D_2'  = \mu_2 I_2 
		\end{array}\right.
	\end{equation}
where $\Nt_i = S_i+I_i+R_i+D_i$ is the constant population size in group $i$.

Throughout this section we will denote solutions related to the classical SIRD model \eqref{Eq.SIRDc} with an index $c$ ($S_{1,c}(t),S_{2,c}(t),I_{1,c}(t), \ldots$) and solutions related to the SIRD model with active population \eqref{Eq.SIRDm} with an index $a$ ($S_{1,a}(t),S_{2,a}(t),I_{1,a}(t), \ldots$).

\begin{example}
As already shown in \cite{Alutto22,Alutto25}, the multigroup model can exhibit multiple waves of infection, which is a phenomenon that does not occur in the classical single group SIR model. Here we present an example that even in the two-group model, we can have 3 waves of infection in the total number of infectious individuals. 

For the set of parameters $\beta_{11} = 0.55$, $\beta_{12} = 15$, $\beta_{21} =0.00002$, $\beta_{22} = 0.55$, $\gamma_1=0.27$, $\gamma_2=0.25$, $\mu_1=\mu_2=0.1$, and initial conditions $N_1(0)=N_2(0)=10^5$, $I_1(0)=500$, $I_2(0)=R_1(0)=R_2(0)=D_1(0)=D_2(0)=0$, $S_1(0)=N_1(0)-I_1(0)$, $S_2(0)=N_2(0)$, the figure shows the functions $I_1(t)$, $I_2(t)$ and $I_1(t)+I_2(t)$.   In each panel, dashed lines correspond to the classical SIRD system \eqref{Eq.SIRDc} and solid lines to the active-population model \eqref{Eq.SIRDm}. Colors encode the group: blue for Group 1, red for Group 2, and black for the total $I_1+I_2$.

In the classical SIRD model, the peak number of infectious individuals occurs during the second wave of infection and is primarily driven by infectious individuals in Group 1. In contrast, in the active population system \eqref{Eq.SIRDm}, the highest number of infectious individuals occurs during the third wave and is mainly caused by infectious individuals in Group 2.

Although the results in \cite{Alutto22,Alutto25} contribute to the understanding of multiple waves of infection in a multigroup model, the full characterization is still not known.

\begin{figure}[H]
	\begin{subfigure}{0.85\textwidth}
		\caption{Infected individuals in Group 1}
		\includegraphics[width=\linewidth]{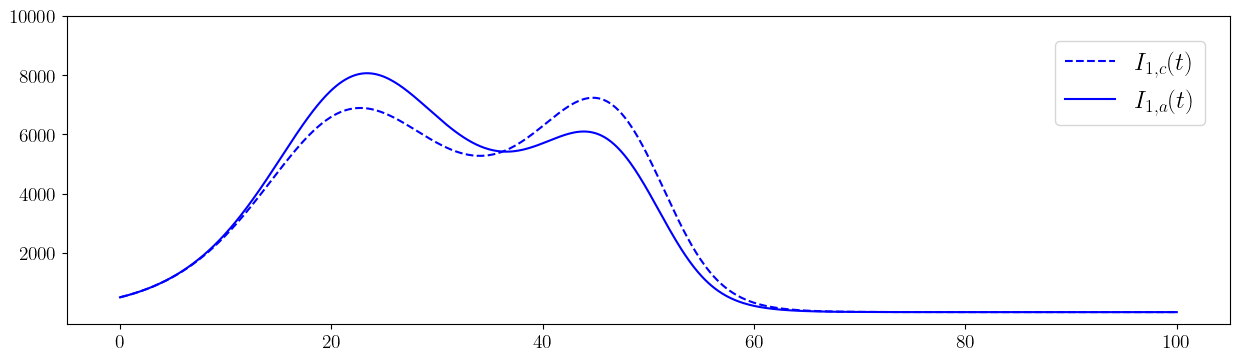}  
		\label{fig.sub.Group1}
	\end{subfigure}
	
	\begin{subfigure}{0.85\textwidth}
		\caption{Infected individuals in Group 2}
		\includegraphics[width=\linewidth]{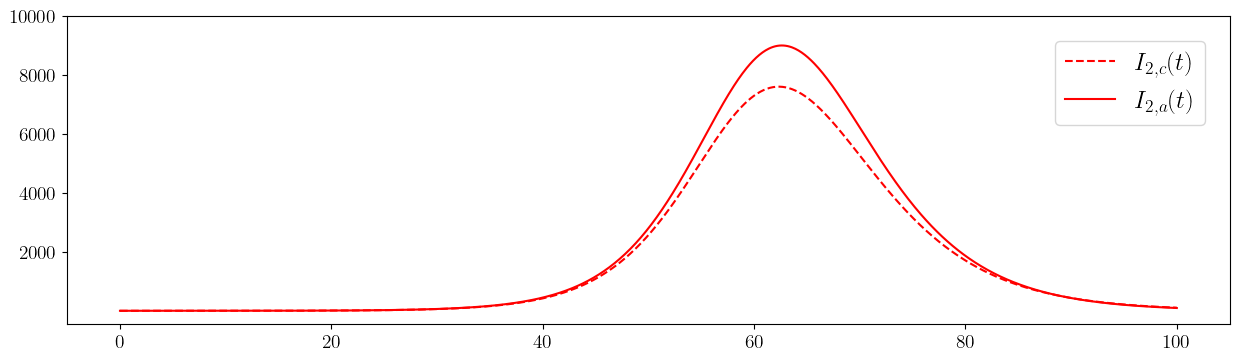}  
		\label{fig.sub.Group2}
	\end{subfigure}
	
	\begin{subfigure}{0.85\textwidth}
		\caption{Total of infected individuals in both groups}
		\includegraphics[width=\linewidth]{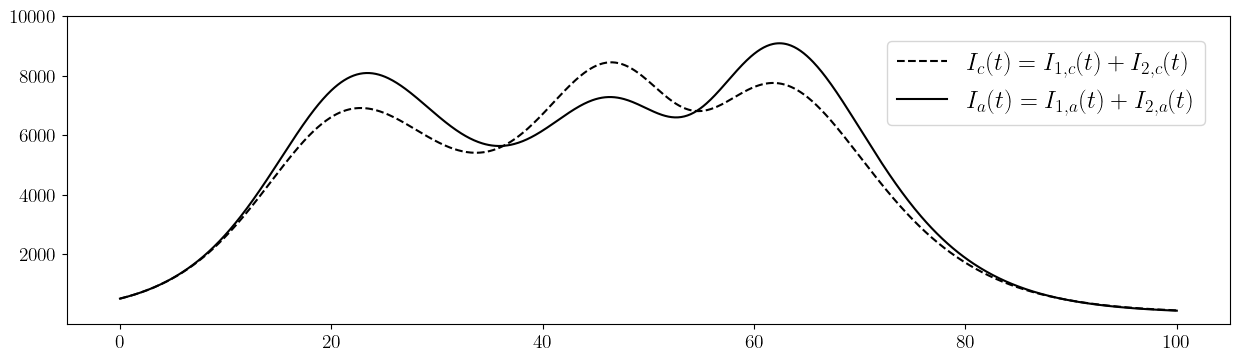}  
		\label{fig.sub.Total}
	\end{subfigure}
	\vspace{-1.5\baselineskip}
    
	\label{Fig.Exemplos}
\end{figure}
\end{example}

\begin{example}\label{ex.S-infty}
Figure \ref{fig.FinalSizeComparison} shows how the final number of susceptible individuals in each group depends on the common mortality rate $\mu$. We denote $S_{i,c}^\infty$ the final state for group $i$ for the classical SIRD model and $S_{i,a}^\infty$ for the active–population model.  

Here we fix $\beta_{11}=0.22, \beta_{12}=0.1, \beta_{21}=0.1, \beta_{22}=0.14$, $\gamma_1=\gamma_2=1/5$, $N_1(0)=N_2(0)=10^5$, $I_1(0)=50$, $I_2(0)=0$, $R_1(0)=R_2(0)=D_1(0)=D_2(0)=0$ e $S(0)=N(0)-I(0)$, and vary $\mu_1=\mu_2=\mu$ in $[0,0.1]$. For each $\mu$, the fixed points $S_{a}^\infty(\mu)$ and $S_{c}^{\infty}(\mu)$ were approximated by iteration of the map $T$ until the tolerance of $10^{-6}$ between the iterations was achieved. For the active population mode we used the map $T$ defined in \eqref{def.T}, and for the classical model we used the equivalent map defined in \cite[p. 2048]{Magal1}.

The horizontal axis represents the common mortality rate $\mu$, and the vertical axis represents the final susceptible level $S_i^\infty$ in group $i$. For each SIRD model, two curves are plotted: blue (group 1) and red (group 2). For each color, the solid line represents the final state $S_{i,a}^\infty$ from the active–population model, while the dashed line represents the final state $S_{i,c}^\infty$ from the classical SIRD model. Comparing the solid and dashed curves of the same color quantifies the impact of the active–population formulation on final epidemic state for that group, for each value of $\mu$.

\begin{figure}[H]
		\caption{Comparison of $S_{i,c}^\infty$ and $S_{i,a}^\infty$ vs $\mu$ (classical vs active)}
		\includegraphics[width=0.9\linewidth]{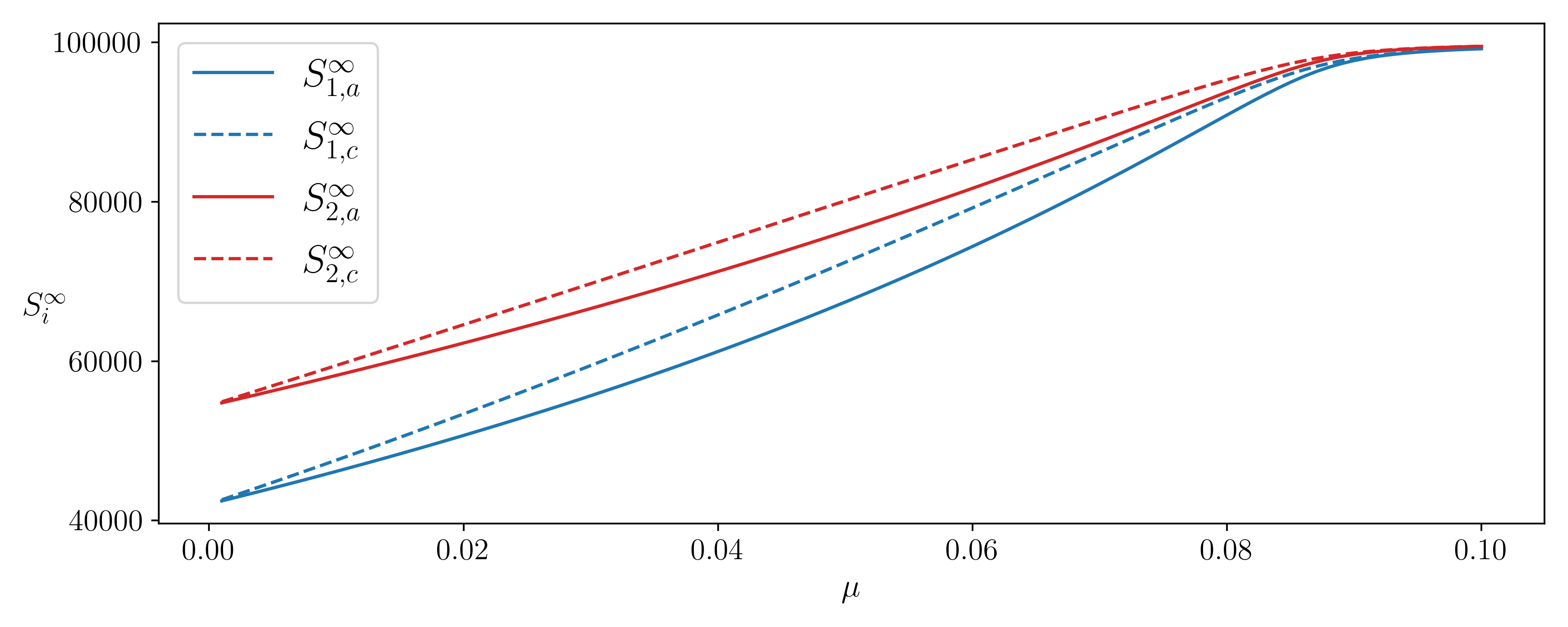}  
		\label{fig.FinalSizeComparison}
\end{figure}
\end{example}

\begin{example}
Here we present the solutions for the susceptible class of active population SIRD system \eqref{Eq.SIRDm} using the same parameters and initial conditions of Example \ref{ex.S-infty}. 

Figure \ref{Fig.Exemplos3} shows: \ref{fig.sub.S1S2truncated}) the evolution of solutions $S_i(t)$ of each group $i$ towards their limit values $S_i^\infty$, which are symbolized by dashed horizontal lines; \ref{fig.sub.Iterations}) the approximation of the limit values $S_i^\infty$, by two methods: the dotted lines with circle markers show the iterations $T^n(0)$, illustrating Theorem \ref{Theo.FinalSize}, converging to the fixed point of $T$; and the solid lines with square markers show the iterations of Newton's Method applied to the function $F(X)=T(X)-X$, so each square represents the coordinates of $X^n=(x_1^n,x_2^n)$ for the sequence $(X^n)_n $ obtained for the approximation for the root of $F$. The initial condition for both methods is $(0,0)$. The horizontal axis in \ref{fig.sub.Iterations} shows the number of iterations. The graph in \ref{fig.sub.S1S2truncated} and the iterations in \ref{fig.sub.Iterations} stop when the tolerance of $10^{-3}$ is achieved.
\begin{figure}[H]
\caption{}
\label{Fig.Exemplos3}
	\begin{subfigure}{0.85\textwidth}
		\caption{$S_i(t)$ e $S_i^\infty$}
		\includegraphics[width=\linewidth]{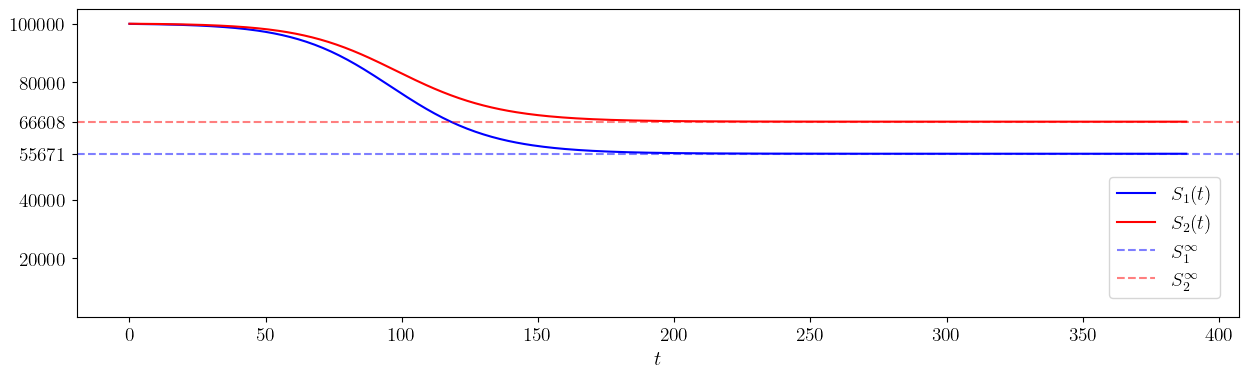}  \label{fig.sub.S1S2truncated}
	\end{subfigure}
    \vspace{-0.5cm}
	
	\begin{subfigure}{0.85\textwidth}
		\caption{Iterations}
        \includegraphics[width=\linewidth]{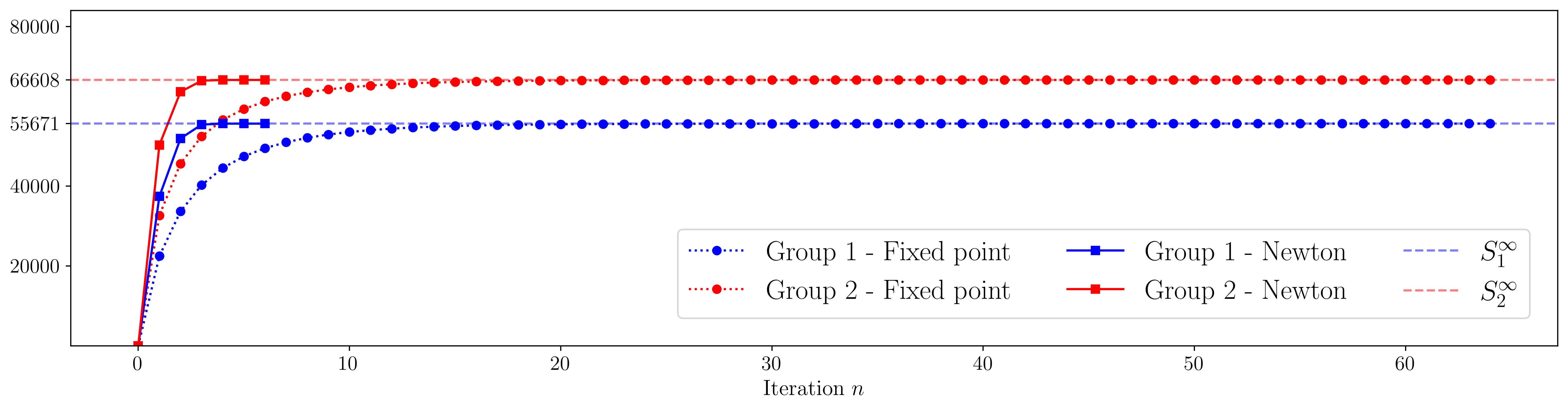} \label{fig.sub.Iterations}
	\end{subfigure}
\end{figure}    
\end{example}

\section{Conclusions}

We studied a two-group SIRD model in which the force of infection is normalized by the active (alive) population $N_i(t)=S_i(t)+I_i(t)+R_i(t)$, rather than by the constant total size $\Nt_i$. This modeling choice introduces an additional nonlinearity through the time dependence of $N_i(t)$ and leads to qualitative and quantitative differences when disease-induced mortality is non-negligible.

A central theoretical outcome is that the final susceptible levels are always strictly positive and can be computed without integrating the ODE system. More precisely, by combining a logarithmic first integral (Theorem~\ref{Thm.PositivityS}) with the algebraic relations between the limit compartments (Lemma~\ref{Lemma.FinalSize1}), we derive an explicit closed final-size system and show that $S^\infty=(S_1^\infty,S_2^\infty)\gg 0$ is a fixed point of a monotone map $T:\mathbb{R}_+^2\to\mathbb{R}_+^2$ (Theorem~\ref{thm:general-final}). The map $T$ depends only on the initial conditions and model parameters (through $S_{i0}$, $N_i(0)$, $R_{i0}$, and the coefficients $\beta_{ij},\gamma_i,\mu_i$), and therefore provides a direct route to the final size that bypasses the full time-dependent dynamics. Under the alternative parameter regimes H2a-H2d, this fixed point is unique and can be obtained by iterating $T$ in $\mathbb{R}^2$ (Theorem~\ref{Theo.FinalSize}), in particular from the simple initial guess $0$, yielding a numerically efficient procedure for approximating $S^\infty$ (and, in the contraction case H2d, with a global linear convergence guarantee).

The fixed point formulation allows us to obtain analytical properties of the final state without using the solution to the ODE system. We illustrated this by proving that the final susceptible state is monotone with respect to transmission: increasing any $\beta_{ij}$ can only decrease $S^\infty$ (Proposition~\ref{prop:monotonicity-order}). Also, we connected early- and late-epidemic descriptors by relating $\mathcal R_0=\rho(\K)$ to $S^\infty$ through a weighted logarithmic identity involving the Perron--Frobenius eigenvector of $\B(\Gamma+\M)^{-1}$ (Theorem~\ref{thm:R0-finalsize}).

Simulations reveal substantial discrepancies between the active population and classical models in transient dynamics and final-size predictions, including scenarios with multiple infection waves and shifts in which group drives the largest peak. We also quantified how varying the common mortality rate $\mu$ affects $S_i^\infty$ and amplifies differences between the two formulations, and illustrated the convergence of the fixed-point approach (alongside Newton-type iterations) as a practical computational tool.

\newpage
\bibliographystyle{siam}
\bibliography{references}

@article{Bui,
author = {Bui, C. and Bethmont, A. and Chughtai, A. A. and Gardner, L. and Sarkar, S. and Hassan, S. and Seale, H. and MacIntyre, C. R.},
title = {A Systematic Review of the Comparative Epidemiology of Avian and Human Influenza A H5N1 and H7N9 – Lessons and Unanswered Questions},
journal = {Transboundary and Emerging Diseases},
volume = {63},
number = {6},
pages = {602-620},
doi = {https://doi.org/10.1111/tbed.12327},
url = {https://onlinelibrary.wiley.com/doi/abs/10.1111/tbed.12327},
eprint = {https://onlinelibrary.wiley.com/doi/pdf/10.1111/tbed.12327},
year = {2016}
}

@article{Nikolay,
author = {Birgit Nikolay  and Henrik Salje  and M. Jahangir Hossain  and A.K.M. Dawlat Khan  and Hossain M.S. Sazzad  and Mahmudur Rahman  and Peter Daszak  and Ute Ströher  and Juliet R.C. Pulliam  and A. Marm Kilpatrick  and Stuart T. Nichol  and John D. Klena  and Sharmin Sultana  and Sayma Afroj  and Stephen P. Luby  and Simon Cauchemez  and Emily S. Gurley },
title = {Transmission of Nipah Virus — 14 Years of Investigations in Bangladesh},
journal = {New England Journal of Medicine},
volume = {380},
number = {19},
pages = {1804-1814},
year = {2019},
doi = {10.1056/NEJMoa1805376},

URL = {https://www.nejm.org/doi/full/10.1056/NEJMoa1805376},
eprint = {https://www.nejm.org/doi/pdf/10.1056/NEJMoa1805376}
}

@article {GaleNikaido,
    AUTHOR = {Gale, David and Nikaid\^o, Hukukane},
     TITLE = {The {J}acobian matrix and global univalence of mappings},
   JOURNAL = {Math. Ann.},
  FJOURNAL = {Mathematische Annalen},
    VOLUME = {159},
      YEAR = {1965},
     PAGES = {81--93},
      ISSN = {0025-5831,1432-1807},
   MRCLASS = {26.55 (90.10)},
  MRNUMBER = {204592},
MRREVIEWER = {V.\ Ganapathy Iyer},
       DOI = {10.1007/BF01360282},
       URL = {https://doi.org/10.1007/BF01360282},
}

@article {Magal1,
	AUTHOR = {Magal, Pierre and Seydi, Ousmane and Webb, Glenn},
	TITLE = {Final size of an epidemic for a two-group {SIR} model},
	JOURNAL = {SIAM J. Appl. Math.},
	FJOURNAL = {SIAM Journal on Applied Mathematics},
	VOLUME = {76},
	YEAR = {2016},
	NUMBER = {5},
	PAGES = {2042--2059},
	ISSN = {0036-1399},
	MRCLASS = {92D30 (92D25)},
	MRNUMBER = {3561786},
	MRREVIEWER = {Yuming Chen},
}

@article {Magal2,
	AUTHOR = {Magal, Pierre and Seydi, Ousmane and Webb, Glenn},
	TITLE = {Final size of a multi-group {SIR} epidemic model: irreducible
		and non-irreducible modes of transmission},
	JOURNAL = {Math. Biosci.},
	FJOURNAL = {Mathematical Biosciences},
	VOLUME = {301},
	YEAR = {2018},
	PAGES = {59--67},
	ISSN = {0025-5564},
	MRCLASS = {92D30},
	MRNUMBER = {3808354},
}

@article {Andreasen,
	AUTHOR = {Andreasen, Viggo},
	TITLE = {The final size of an epidemic and its relation to the basic
		reproduction number},
	JOURNAL = {Bull. Math. Biol.},
	FJOURNAL = {Bulletin of Mathematical Biology. A Journal Devoted to
		Research at the Interface of the Life and Mathematical
		Sciences},
	VOLUME = {73},
	YEAR = {2011},
	NUMBER = {10},
	PAGES = {2305--2321},
	ISSN = {0092-8240},
}

@article{NGM,
	author = {Diekmann, O.  and Heesterbeek, J. A. P.  and Roberts, M. G. },
	title = {The construction of next-generation matrices for compartmental epidemic models},
	journal = {Journal of The Royal Society Interface},
	volume = {7},
	number = {47},
	pages = {873-885},
	year = {2010},
}

@article{KmcK,
	author = {Kermack, William Ogilvy  and McKendrick, A. G.  and Walker, Gilbert Thomas },
	title = {A contribution to the mathematical theory of epidemics},
	journal = {Proceedings of the Royal Society of London. Series A, Containing Papers of a Mathematical and Physical Character},
	volume = {115},
	number = {772},
	pages = {700-721},
	year = {1927}
}

@article {Miller,
    AUTHOR = {Miller, Joel C.},
     TITLE = {A note on the derivation of epidemic final sizes},
   JOURNAL = {Bull. Math. Biol.},
  FJOURNAL = {Bulletin of Mathematical Biology. A Journal Devoted to
              Research at the Interface of the Life and Mathematical
              Sciences},
    VOLUME = {74},
      YEAR = {2012},
    NUMBER = {9},
     PAGES = {2125--2141},
      ISSN = {0092-8240,1522-9602},
   MRCLASS = {92D30},
  MRNUMBER = {2964890},
MRREVIEWER = {Andrea\ Velenich},
       DOI = {10.1007/s11538-012-9749-6},
       URL = {https://doi.org/10.1007/s11538-012-9749-6},
}

@article {MKS,
    AUTHOR = {Mukherjee, Abhik and Kundu, Souvik and Sasmal, Sourav Kumar},
     TITLE = {Final size relations for some compartmental models in
              epidemiology},
   JOURNAL = {J. Biol. Systems},
  FJOURNAL = {Journal of Biological Systems},
    VOLUME = {32},
      YEAR = {2024},
    NUMBER = {2},
     PAGES = {921--940},
      ISSN = {0218-3390,1793-6470},
   MRCLASS = {92D30},
  MRNUMBER = {4750739},
       DOI = {10.1142/S0218339024500311},
       URL = {https://doi.org/10.1142/S0218339024500311},
}

@article {Arino,
    AUTHOR = {Arino, Julien and Brauer, Fred and van den Driessche, P. and
              Watmough, James and Wu, Jianhong},
     TITLE = {A final size relation for epidemic models},
   JOURNAL = {Math. Biosci. Eng.},
  FJOURNAL = {Mathematical Biosciences and Engineering. MBE},
    VOLUME = {4},
      YEAR = {2007},
    NUMBER = {2},
     PAGES = {159--175},
      ISSN = {1547-1063,1551-0018},
   MRCLASS = {92D30 (34C11 34D20)},
  MRNUMBER = {2293759},
       DOI = {10.3934/mbe.2007.4.159},
       URL = {https://doi.org/10.3934/mbe.2007.4.159},
}

@article {MaEarn,
    AUTHOR = {Ma, Junling and Earn, David J. D.},
     TITLE = {Generality of the final size formula for an epidemic of a
              newly invading infectious disease},
   JOURNAL = {Bull. Math. Biol.},
  FJOURNAL = {Bulletin of Mathematical Biology. A Journal Devoted to
              Research at the Interface of the Life and Mathematical
              Sciences},
    VOLUME = {68},
      YEAR = {2006},
    NUMBER = {3},
     PAGES = {679--702},
      ISSN = {0092-8240,1522-9602},
   MRCLASS = {92D30 (34C11)},
  MRNUMBER = {2224786},
       DOI = {10.1007/s11538-005-9047-7},
       URL = {https://doi.org/10.1007/s11538-005-9047-7},
}

@INPROCEEDINGS{Alutto22,
  author={Alutto, Martina and Cianfanelli, Leonardo and Como, Giacomo and Fagnani, Fabio},
  booktitle={2022 IEEE 61st Conference on Decision and Control (CDC)}, 
  title={Multiple peaks in network SIR epidemic models}, 
  year={2022},
  volume={},
  number={},
  pages={5614-5619},
  doi={10.1109/CDC51059.2022.9992408}}

@ARTICLE{Alutto25,
  author={Alutto, Martina and Cianfanelli, Leonardo and Como, Giacomo and Fagnani, Fabio},
  journal={IEEE Transactions on Control of Network Systems}, 
  title={On the Dynamic Behavior of the Network SIR Epidemic Model}, 
  year={2025},
  volume={12},
  number={1},
  pages={177-189},
  doi={10.1109/TCNS.2024.3448136}}

@article{Li2020_COVID,
  author  = {Li, Q. and Guan, X. and Wu, P. and Wang, X. and Zhou, L. and Tong, Y. and others},
  title   = {Early Transmission Dynamics in Wuhan, China, of Novel Coronavirus–Infected Pneumonia},
  journal = {New England Journal of Medicine},
  year    = {2020},
  volume  = {382},
  number  = {13},
  pages   = {1199--1207},
  doi     = {10.1056/NEJMoa2001316}
}

@article{Verity2020_IFR,
  author  = {Verity, R. and Okell, L. C. and Dorigatti, I. and Winskill, P. and Whittaker, C. and Imai, N. and others},
  title   = {Estimates of the severity of coronavirus disease 2019: a model-based analysis},
  journal = {The Lancet Infectious Diseases},
  year    = {2020},
  volume  = {20},
  number  = {6},
  pages   = {669--677},
  doi     = {10.1016/S1473-3099(20)30243-7}
}

@article{WHO2014_Ebola,
  author  = {{WHO Ebola Response Team}},
  title   = {Ebola Virus Disease in West Africa -- The First 9 Months of the Epidemic and Forward Projections},
  journal = {New England Journal of Medicine},
  year    = {2014},
  volume  = {371},
  number  = {16},
  pages   = {1481--1495},
  doi     = {10.1056/NEJMoa1411100}
}

@misc{WHO_Nipah_2023,
  author       = {{World Health Organization}},
  title        = {Nipah virus},
  year         = {2023},
  howpublished = {\url{https://www.who.int/news-room/fact-sheets/detail/nipah-virus}},
  note         = {Acessado em janeiro de 2026}
}

@article{Lai2016_H5N1,
  author  = {Lai, Shengjie and Qin, Yuntao and Cowling, Benjamin J. and others},
  title   = {Global epidemiology of human infections with avian influenza A(H5N1) virus, 1997--2015: a systematic review},
  journal = {Journal of Infection},
  year    = {2016},
  volume  = {72},
  number  = {6},
  pages   = {70--87},
  doi     = {10.1016/j.jinf.2015.10.002}
}

@article{Aditama2012_H5N1,
  author  = {Tjandra Yoga Aditama and Vivek Shinde and Michael W. Lindblade and
             Timothy Uyeki and Wiku Adisasmito and Umar Fahmi Achmadi and
             John A. Tjandrawinarso and {et al.}},
  title   = {Avian Influenza H5N1 Transmission in Households, Indonesia},
  journal = {PLoS ONE},
  year    = {2012},
  volume  = {7},
  number  = {1},
  pages   = {e29971},
  doi     = {10.1371/journal.pone.0029971},
  url     = {https://journals.plos.org/plosone/article?id=10.1371/journal.pone.0029971}
}

@article{Park2018_Transmissibility,
  author  = {Jong Eun Park and Kyungmin Ryu and Ji Yun Noh and Woo Joo Kim and
             Hee Jin Cheong},
  title   = {Transmissibility and severity of influenza virus by subtype},
  journal = {Infection, Genetics and Evolution},
  year    = {2018},
  volume  = {65},
  pages   = {288--292},
  doi     = {10.1016/j.meegid.2018.08.009},
  url     = {https://www.sciencedirect.com/science/article/pii/S1567134818306051}
}

\end{document}